\documentclass[]{elsart3p}
 \usepackage{graphicx}
 \usepackage{epsfig}

\usepackage{amssymb}

\journal{Nucl. Instr. and Meth. B}

\begin{document}
\begin{frontmatter}
\title{Proton-induced magnetic order in carbon: SQUID measurements\thanksref{label1}}
\author{J. Barzola-Quiquia,}
\author{S. Petriconi,}
\author{P. Esquinazi\corauthref{cor1}},
\ead{esquin@physik.uni-leipzig.de}
\author{M. Rothermel,}
\corauth[cor1]{Corresponding author}
\author{D. Spemann,}
 \author{A. Setzer,}
\author{T. Butz}
\thanks[label1]{This work was supported by the DFG under grant DFG ES 86/11-1
and by the European Union under "Ferrocarbon". Discussions with R.
H\"ohne and Y. Kopelevich are gratefully acknowledged. We are in debt
with D. Alber from the Hahn-Meitner-Institut in Berlin, where the NAA
analysis at BER II were done, as well as with T. Agn\'e for arranging
this measurement. We thank P. Morgenstern from the Institute for
Surface Modification (IOM) in Leipzig for the XRF measurements.}
\address{Institut f\"ur Experimentelle Physik II, Universit\"at Leipzig,
Linn\'estrasse 5, D-04103 Leipzig, Germany}




\begin{abstract}
In this work we have studied systematically the changes in the
magnetic behavior of highly oriented pyrolytic graphite (HOPG)
samples after proton irradiation in the MeV energy range.
Superconducting quantum interferometer device (SQUID) results
obtained from samples with thousands of localized spots of micrometer
size as well on samples irradiated with a broad beam confirm
previously reported results. Both, the para- and ferromagnetic
contributions depend strongly on the irradiation details. The results
indicate that the magnetic moment at saturation of spots of
micrometer size is of the order of $10^{-10}~$emu.
\end{abstract}

\end{frontmatter}

\section{Introduction}
\label{i} Magnetic order at room temperature in metal-free
carbon-based structures remains one of the exciting issues in
fundamental and applied research across all scientific disciplines.
However, the lack of reproducibility of early results added to the
unknown, in some cases late characterization of the magnetic
impurities \cite{retraction} increased substantially the scepticism
of the scientific community. In the year 2003 some of us reported
that proton irradiation of MeV energy on HOPG samples triggers ferro-
or ferrimagnetism at room temperature \cite{pabloprl03,han03}.
Although reports on the reproducibility of  this behavior in two
different HOPG samples followed the original publication
\cite{esq05}, the lack of independently published studies that
successfully triggered magnetic order after ion irradiation of carbon
pushed us to elaborate and extend our experimental work.

Although there are several theoretical works in the literature on
magnetic order in carbon, specially hydrogen-induced (see for example
Ref.~\cite{ma05} and references therein), apart from the proton
induced magnetic order in graphite there are no much new systematic
experimental works that show ferromagnetism in metal-free carbon,
including in particular a rigorous characterization of the samples
impurities. The study done in Ref.~\cite{tala05} reported that
Nitrogen and Carbon irradiation of nanosized diamond powder triggers
magnetic order at room temperature. However, no study of the impurity
concentration was presented in that work.  The studies done in
\cite{sai05} reveal that carbon films prepared by CVD on stainless
steel substrates reach magnetization values of the order of
0.15~emu/g at room temperature, comparable to those reported in
earlier studies \cite{murata91,murata92}. In that work the amount of
measured impurities appears to be not enough to account for the
absolute value of the magnetic moments of the samples. In
Ref.~\cite{lee05} the magnetization of proton irradiated graphite was
studied after one low-dose irradiation. The authors show there that
proton irradiation induces Curie-type paramagnetism in graphite but
no measurements of the hysteresis loops before and after irradiation
were apparently done in order to check for the induced magnetic
order. The experimental difficulties to reproduce this magnetic order
and the weakness of the ferromagnetic signals, which are sometimes at
the limit of the sensitivity of current experimental characterization
methods, including SQUID's, are the main obstacles that preclude a
rush development of this interesting and important subject.

The aim of this report is threefold. Firstly, we discuss the expected
ferromagnetic signals after proton irradiation based on our previous
publication \cite{pabloprl03} and the necessary precautions one needs
to take for the SQUID measurements of these weak signals. To increase
the relative sensitivity of these measurements and to avoid
experimental artifacts as well as any introduction of magnetic
impurities through sample handling, we developed a new, simple sample
holder that allows consecutive irradiation and SQUID measurements
without touching the sample or its holder as well as their relative
positions. Secondly, in order to check the magnetic impurity
concentration of the HOPG samples and the accuracy of the method we
used (Particle Induced X-ray Emission (PIXE)), we have compared the
impurity concentrations measured with three experimental methods on
similar HOPG samples. Thirdly, we have irradiated three HOPG samples
under different conditions and measured the changes in their magnetic
behavior with a SQUID. We show that proton irradiation induces
ferromagnetism in graphite, as demonstrated earlier\cite{pabloprl03},
and we extend our studies to the induced paramagnetism. Through the
irradiation of thousands of spots of micrometer size in a single
sample we were able to measure their ferromagnetic signal and
estimate the average magnetic moment of each spot.


\section{Experimental details}
\label{e}

The studies performed in Ref.~\cite{pabloprl03} indicate that the
ferromagnetic magnetic moment at saturation $m_s$ depends on the
total implanted charge $C_t$. Roughly speaking  $m_s \sim 2 \times
10^{-7}$~[emu/$\mu$C$^{0.5}] C_t^{0.5}$ for $C_t < 10^3~\mu$C (the
usual magnetic moment (cgs) unit ``emu" is equal to $10^{-3}$Am$^2$).
 $C_t$ means the total charge irradiated on the same area. The
square root dependence suggests that at small doses the experimental
signal should be directly proportional to the irradiation dose, which
produces vacancies, adatom defects and/or vacancy-hydrogen complexes.
However, it tends to saturate at higher doses, probably due to damage
accumulation (e.g. saturation of dangling bonds), hydrogen outgassing
and/or annealing of certain defects \cite{lehtinen04}. This square
root relation should be taken only as a rough estimate for the
expected ferromagnetic signal and for broad irradiation areas. The
actual value of $m_s$, however, depends on parameters like proton
current, i.e. the higher the current the lower is $m_s$, (e.g. due to
heating effects), or beam size (broad or narrow beam irradiation)
\cite{esq05}. According to earlier results \cite{pabloprl03} we
expect $m_s$ of the order of $10^{-6}$~emu for a total irradiated
charge of $100~\mu$C. That means that the SQUID magnetometer should
provide reliable and reproducible results within a magnetic moment
range better than $1~\mu$emu, specially after introducing and taking
out the sample with holder several times into or from the SQUID.

Usually at fields $B \lesssim 1~$T the reproducibility of commercial
SQUID's is better than $1~\mu$emu and therefore these magnetometers
can be used to measure the effects produced by irradiation (an
example is shown below). Care should be taken, however, with possible
artifacts of these systems, specially magnetic field hysteresis due
to the electronics and/or superconducting solenoid properties
\cite{rh}. The reproducibility of each SQUID system should be checked
before starting the irradiation steps.

\begin{figure}
\centerline{\psfig{file=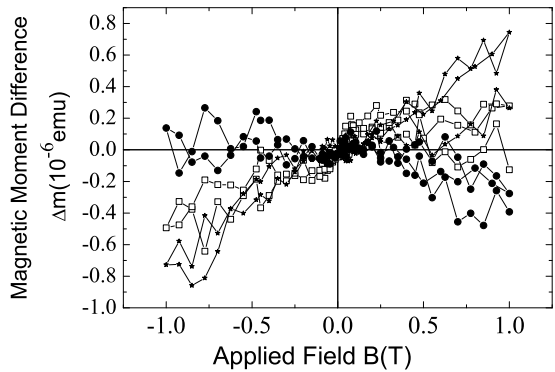,width=8.5cm}} \caption{Difference of
magnetic moments as a function of field (hysteresis loops) measured
for the same HOPG sample and holder (in this case number~3 as
example) at four different days at $T = 100~$K. The difference
$\Delta m$ is calculated taking the measurement at a certain day as
reference. Similar differences are obtained at all temperatures and
choosing measurements at other days as reference.} \label{back}
\end{figure}

The magnetic moment measurements were performed with a SQUID
magnetometer  from Quantum Design with the reciprocating sample
option (RSO). We note that the SQUID sensitivity without this option
is not enough to measure accurately the effects produced by
irradiation, specially when the magnetic signal is of the order of
$\mu$emu for the saturation ferromagnetic moment. The magnetic field
was applied parallel to the graphene planes in all measurements in
order to diminish the diamagnetic background. As an example for the
reproducibility and error, Fig.~\ref{back} shows the difference
between hysteresis loops of the same HOPG sample (including holder)
before irradiation measured at different days. Each measurement was
performed following the same sequence after removing and introducing
the whole sample with holder from/into the SQUID apparatus, leaving
the sample-holder several days at ambient conditions between the
measurements.
The difference $\Delta m$ plotted in Fig.~\ref{back} should be
ideally zero at all fields. We recognize, however, that both
paramagnetic- as diamagnetic-like deviations are obtained after
subtraction of an arbitrarily chosen hysteresis loop (measured at a
certain day) from the other loops. The largest deviation obtained is
of the order of $8 \times 10^{-7}~$emu at a field of 1~T. Because
most of the HOPG samples we have measured in their virgin states show
ferromagnetic-like hysteresis loops, which origin is not related to
the magnetic impurities \cite{pabloprb02}, the effects produced by
the irradiation are much better identified doing this point-by-point
(at equal ($B,T$)) difference in $\Delta m$.

Because we do not know the amount of sample that remains magnetic
after irradiation, all the SQUID data are presented as magnetic
moment directly measured with this method. Although one knows the
penetration depth of protons of the used energy in graphite
($\lesssim 48~\mu$m) it is still unclear, which is the range where
the main ferromagnetic and paramagnetic signals come from. Therefore,
it has little sense to divide the measured values of magnetic moment
by the total sample mass. Taking into account recently done x-ray
magnetic circular dichroism and magnetic force microscopy
measurements of irradiated spots in 200~nm carbon films, which
provide clear evidence for the existence of magnetic order at the
spot position \cite{ohldag}, we tend to assume that the ferromagnetic
layer in our samples should be mainly located at the first micrometer
from the sample surface.

The irradiations were done with the high-energy nanoprobe LIPSION, a
single ended 3MV SINGLETRON  accelerator with a RF-source for protons
and alpha particles. Two irradiation chambers allow the irradiation
with proton beams of diameters as low as 50~nm up to 0.8~mm at
different proton currents at MeV energy.

All three samples were HOPG from Advanced Ceramics with a rocking
curve width of $0.4^\circ$ (grade A).\\
\noindent -(a) Sample~1 had a mass of 3.5~mg with a size $2 \times 3
\times 0.3~$mm$^3$. As in Ref.~\cite{pabloprl03}, it was glued on a
high purity Si substrate with varnish. The irradiation of this sample
consisted on three broad proton spots of 2~MeV proton energy, 0.8~mm
diameter each, 0.7~mm
distance, $150~\mu$C total charge per spot and a proton current of 100~nA.\\
-(b) Sample~2 of 5.7~mg weight and $2 \times 4 \times 0.3~$mm$^3$
size was glued in the middle of a 13.5~cm long and 1~mm diameter pure
Cu wire. Due to the long length and the homogeneous magnetic response
of the wire, it provides a small contribution to the SQUID signal.
The whole ensemble sample plus wire including the corresponding
couplings for the SQUID and accelerator chamber, were used for the
irradiation of: first $10^4$~spots and later $2 \times 10^4$ spots
more. Each spot was made with a 2.25 MeV proton energy, beam diameter
of $\simeq 2~\mu$m, a total charge of $1.16~$nC (fluence
$0.37~$nC/$\mu$m$^2 = 2.3 \times 10^{17}~$protons/cm$^{2}$)
and we used 700~pA proton current.\\
-(c) Sample~3 had a mass of  10.8~mg and a size of $4 \times 3 \times
0.4$~mm$^3$. It was glued with varnish on the middle of a 13.5~cm
long pure quartz rod. Before fixing it a 20~nm gold film was
deposited on the rod, which is necessary for the measurement of the
proton current in the accelerator chamber during irradiation. For the
irradiation of spots we used a $\simeq 1.5~\mu$m diameter proton
beam, which irradiated a $\simeq 5~\mu$m diameter area following a
computer controlled spiral movement. 400 spots ($25~\mu$m apart) with
a total charge of $191~\mu$C, fluence per spot of
$24.3~$nC/$\mu$m$^2$ ($1.52 \times 10^{19}~$cm$^{-2}$) at a proton
current of 6.5~nA (10 times larger than for sample~2) and 2.25 MeV
energy were prepared in sample~3.

\section{Results}
\subsection{Impurity measurements}
\label{im} As described in section~\ref{i}, we expect
ferromagnetic-like signals of the order of a few $\mu$emu. How much
ferromagnetic Fe is necessary to produce a magnetic moment of, e.g.,
$5~\mu$emu? From literature data one estimates easily that 23~ng pure
Fe (or a volume of $\sim 3 \times 10^{-9}~$cm$^3$) would be enough to
produce this magnetic moment assuming that this small amount is
ferromagnetic at room temperature. Under these assumptions the
relative Fe concentration in a typical HOPG sample would be of the
order of $6~\mu$g/g. That means that we need impurity measurements
that provide a sensitivity of at least $1~\mu$g/g for Fe. The method
called Particle Induced X-ray Emission (PIXE) has this sensitivity
(or better) for the analysis of Fe in a carbon matrix. Our PIXE
measurements performed in situ and during irradiation show that the
total amount of Fe impurities in our HOPG samples is $0.6 \pm
0.04~\mu$g/g (i.e. 0.15~ppm)\cite{esq05} for grade A samples. To
check the accuracy of our PIXE analysis we have measured similar HOPG
samples with two other methods. Neutron Activation Analysis (NAA) on
a 30~mg HOPG grade B sample with an neutron activation time of 18
days and $\gamma$-ray measuring time of 6 hours provides a total Fe
impurities of $0.17 \pm 0.03~\mu$g/g. Our PIXE measurements on the
same sample gave for Fe $0.17~\mu$g/g as well. The third method used
was x-ray fluorescence with a EDXRF-spectrometer Quan X. The Fe
concentration in the same HOPG grade A sample measured with PIXE was
below the minimum detection limit ($\sim 5~\mu$g/g) this analytical
method has. The amount of other magnetic impurities was much below
that for Fe (for example, $2.5 \times 10^{-3}~\mu$g/g for Co).

Concluding, assuming the worst, unlikely case, the maximum
ferromagnetic magnetic moment at saturation one expects from this
amount of magnetic impurities in our HOPG samples -- were this amount
ferromagnetic at room temperature -- is $\lesssim 5 \times
10^{-7}~$emu. Certainly, one should rule out that due to an improper
sample handling and between irradiation steps a small Fe grain with a
mass of a few tens of ng does get fixed somewhere at the surface of
the HOPG sample. Therefore, special holders as well as systematic
irradiation steps are necessary.

\subsection{Broad irradiation}
\label{b}  
\begin{figure}
\centerline{\psfig{file=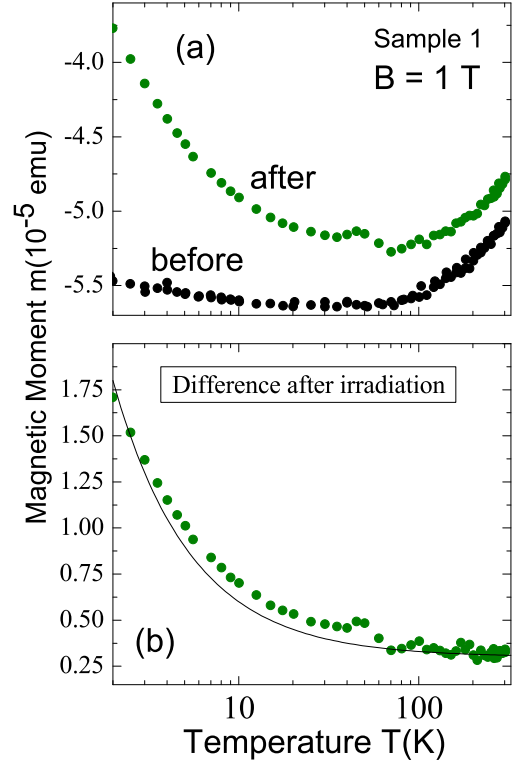,width=8.0cm}} \caption{(a) Total
magnetic moment (HOPG sample~1 with the Si substrate) as a function
of temperature at a constant magnetic field applied parallel to the
graphene planes for the sample before and after proton irradiation
with a broad beam (total charge $450~\mu$C at 100~nA proton current).
(b) The difference between the two curves from (a). This difference
reveals directly the irradiation effect. The continuous line is the
function $3 \times 10^{-5}$[emu K]$/T + 3 \times 10^{-6}$[emu].}
\label{s11}
\end{figure}

The temperature dependence of the magnetic moment before
and after irradiation of sample~1 is shown in Fig.~\ref{s11}(a) in a
semilogarithmic scale. The temperature dependence of the virgin curve
shows a minimum (maximum diamagnetism) at $T \sim 30~$K, which is
usual for HOPG samples of good quality. After a broad proton beam
irradiation covering most of the sample area, the magnetic moment
shows a clear increase in all the temperature range.
Figure~\ref{s11}(b) shows the difference between the magnetic moment
after the irradiation minus that of the virgin state as a function of
temperature at a constant field of 1~T. This difference can be
roughly understood as the sum of two contributions, namely, a
paramagnetic one, which follows roughly the Curie law $3 \times
10^{-5}/T$~emu, and a ferromagnetic constant contribution $3 \times
10^{-6}~$emu, i.e. $m(T,B>B_s) \simeq \frac{3 \times 10^{-5}}{T} + 3
\times 10^{-6}$, where $B_s$ is the minimum saturation field for the
ferromagnetic part. The small but clear deviation between the fit and
the data shown in Fig.~\ref{s11}(b) can be interpreted as follows.
The main variable of the Brillouin function $B_J(x)$ with $x =
gJ\mu_B B/k_B T$ reaches relatively high values at low temperatures
for an applied field of 1~T. Assuming for simplicity the product of
the Land\'e factor $g$ with the total angular momentum $\vec{J}$, $gJ
\sim 1$, this variable is $0.13 \le x \le 2.2 \times 10^{-3}$ for
5~K$ \le T \le$ 300~K at $B = 1~$T. Only for $x \ll 1$ one is allowed
to keep the first term of the Brillouin function that provides the
simple $1/T$ Curie law. Part of the  deviation may also come from the
assumption of a strictly temperature independent ferromagnetic
contribution.

\begin{figure}
\centerline{\psfig{file=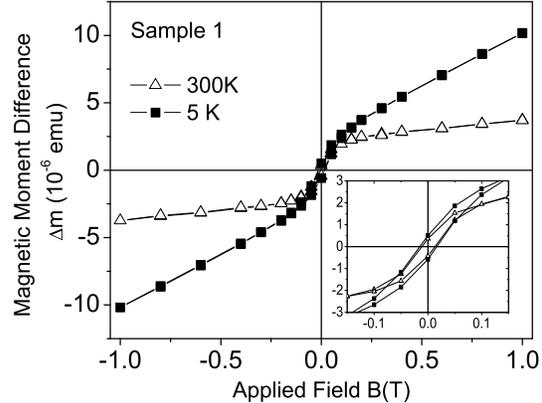,width=8.0cm}}
\caption{Hysteresis loops for sample~1 obtained from the difference
between the loops measured after and before irradiation $\Delta m =
m_a - m_b$ at the same magnetic fields in each state at two
temperatures. The inset blows up the data in a smaller field range.}
\label{s12}
\end{figure}

The hysteresis loops shown in Fig.~\ref{s12}, obtained at two
temperatures subtracting the loops after irradiation from those
obtained in the virgin state, justify the assumption of the two
magnetic contributions. The increase of the magnetic moment after
irradiation at room temperature is mainly due to the increase of the
ferromagnetism of the sample; only $\sim 25\%$ of the increase is due
to a paramagnetic contribution at 300~K and 1~T. The paramagnetic
contribution is clearly recognized in Fig.~\ref{s12} from the slope
of  the loops at fields above $\sim 0.25~$T. The inset in this figure
shows clearly the finite irreversibility produced by the irradiation
with coercivity fields of the order of 0.02~T.

Taking into account the total irradiated charge per spot and after
Ref.~\cite{pabloprl03} we expect a $m_s(300) \sim 3 \times 2 \times
10^{-7} \sqrt{150} \simeq 7 \times 10^{-6}~$emu, in comparison we
obtain $m_s (300) \sim 3.0 \times 10^{-6}~$emu. The results shown in
Fig.~\ref{s11} clearly indicate that broad irradiation -- at the used
proton current and fluence --
 triggers two magnetic contributions, one due to independent, localized magnetic
moments (e.g. dangling bonds due to the disorder produced by the
irradiation) and a second one with all the characteristics of
magnetic order with a Curie temperature  above room temperature. We
stress that the effect on the magnetic properties of graphite due to
proton irradiation depends on se\-veral parameters as  the total
implanted or irradiated charge, fluence and proton current as well as
on the geometry of the used proton beam, as the next section
discusses.

\subsection{Magnetic spots of micrometer size}
\subsubsection{Low proton current} \label{m}
\begin{figure}
\centerline{\psfig{file=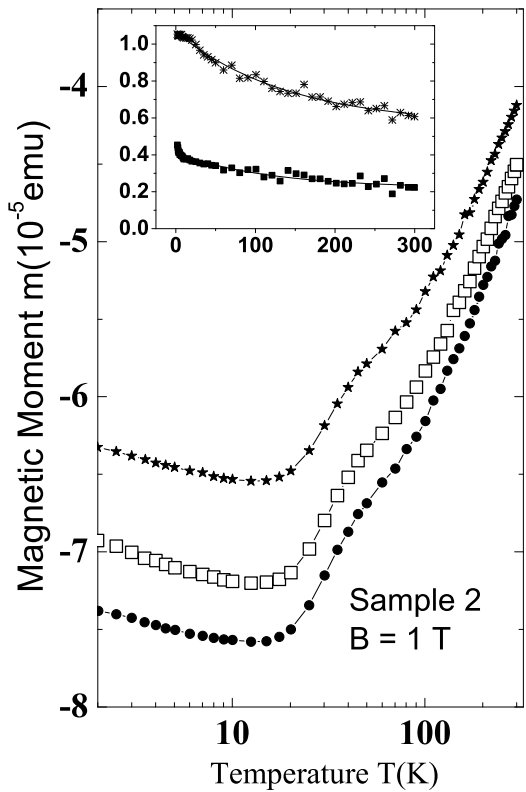,width=8.3cm}} \caption{Total
magnetic moment of sample~2 as a function of temperature at a field
of 1~T for the virgin $(\bullet)$, after the first irradiation with
$10^4$ spots (total charge $11.6~\mu$C at 700~pA ($\square$)), and
after the second irradiation adding $2 \times 10^4$ similar spots as
in the first irradiation $(\bigstar)$. The inset shows the difference
of magnetic moment between the first $(\blacksquare)$ (second
$(\star)$) irradiation and the virgin state. The continuous lines are
fits to the function $m_0 + m_1 \exp(-T/T_0)$ with the parameters
$m_0 = 2.2 (5.6) \times 10^{-6}~$emu, $m_1 = 1.9 (5.07) \times
10^{-6}~$emu and $T_0 = 116 (140)~$K for the first (second)
irradiated sample.} \label{s21}
\end{figure}

Localized proton irradiation of spots of micro\-meter size triggers a
slightly different magnetic response on graphite. Figure~\ref{s21}
shows the total magnetic moment (sample and holder) as a function of
temperature at a field of 1~T for sample 2 in three different states,
virgin, with $10^4$ and with $3 \times 10^4$ irradiated spots, each
of them with a total charge of 1.16~nC. The temperature dependence of
$m$ is quali\-tatively similar to that of Fig.~\ref{s11}. The
difference is mainly due to the sample misalignment respect to the
applied field and partially also to a different contribution of the
sample holder.

\begin{figure}
\centerline{\psfig{file=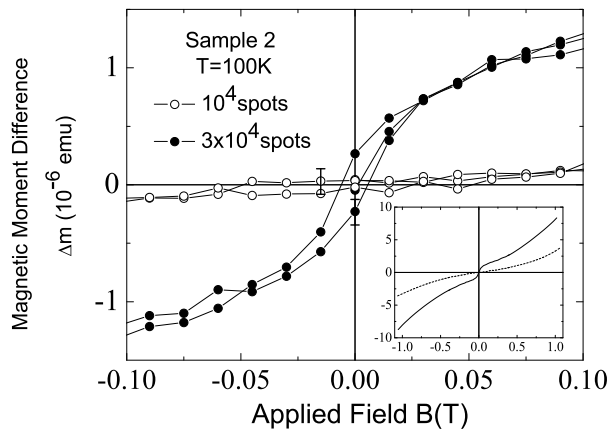,width=9.0cm}}
\caption{Hysteresis loops for sample~2 obtained from the difference
between the loops measured after and before irradiation $\Delta m =
m_a - m_b$ at the same magnetic fields in the states after two
irradiations at 100~K. The points $(\circ)$ were obtained after
irradiation of $10^4$ spots (11.6~$\mu$C total charge at 700~pA
proton current) and $(\bullet)$ adding $2 \times 10^4$ (34.8~$\mu$C
total charge) spots on a different area of the same sample. The bars
indicate the maximum expected error due to the reproducibility of our
SQUID and the subtraction. The inset shows the loops in a broader
field range after the first (dashed line) and second irradiation
(continuous line).} \label{s22}
\end{figure}

The irradiation effect on the temperature dependence can be better
recognized from the difference between the $m(T)$ curves shown in the
inset of Fig.~\ref{s21}. The inset shows
 that no Curie-like $(1/T)$-contribution is
obtained after producing the localized spots. Part of the magnetic
response of the spots is due to ferromagnetism, as the hysteresis
loops shown in Fig.~\ref{s22} indicate. From these loops we obtain a
saturation magnetic moment $m_s \sim 10^{-6}~$emu. For the first
irradiation with only $10^4$ spots the error involved in the
subtraction, specially at low fields, does not permit to assure the
existence of an hysteresis. We stress that signals of the order of $2
\times 10^{-7}$~emu are at the limit of reliability (not resolution!)
of our SQUID apparatus.

It is well known that the paramagnetic Curie law holds only if $x \ll
1$. This law is a consequence of thermal average involving $(2J+1)$
equally spaced levels, which originate from the effect of the applied
field on one multiplet. If at large magnetic fields new multiplet
levels start to contribute to the statistical average or if their
energy levels are not equally spaced, then deviations of $m(T)$ from
the Curie law to a weaker $T-$dependence are expected. The observed
positive curvature for fields above 0.25~T in the hysteretic loops,
see inset in Fig.~\ref{s22}, suggests that the applied field
influences the number of multiplet levels that determine the total
magnetic moment. The clearly weak $T-$dependence of the irradiated
sample 2, see inset in Fig.~\ref{s21}, agrees with this expectation.
The $T-$dependence shown in this inset can be fitted with an
exponential function of the form $m \simeq m_0 + m_1 \exp(-T/T_0)$
with $m_{0,1}$ and $T_0$ free fitting parameters. If we use the
Brillouin function plus a $T-$independent ferromagnetic contribution
we can in principle fit the measured $T-$dependence but using $gJ$
values that are above any reasonable limit, indicating its inadequacy
to understand the magnetism of the irradiated spots. Whatever the
reasons for the observed behavior, the measured curves indicate a
constant temperature ferromagnetic-like term of the order of $m_0
\simeq 2 \times 10^{-6}$~emu for the sample with the first $10^4$
spots and $m_0 \simeq 5.6 \times 10^{-6}~$emu for the sample with $3
\times 10^4~$ spots at 1~T. These numbers indicate a magnetic moment
of the order of $2 \times 10^{-10}~$emu per spot produced at the
conditions described in section~\ref{e}.

\subsubsection{Large proton current}
\begin{figure}
\centerline{\psfig{file=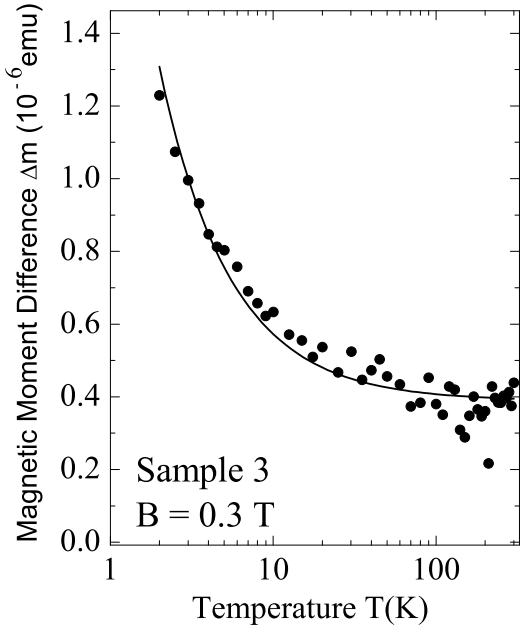,width=8.3cm}}
\caption{Temperature dependence of the difference between the
magnetic moments measured after and before the irradiation of 400
micrometer-size spots with a total charge of 191~$\mu$C at a current
of 6.5~nA in sample~3. The measurements were done at a constant field
of 0.3~T. The continuous line follows the equation $3.9 \times
10^{-7}$~[emu] $+ 1.84 \times 10^{-6}$~[emu K]$/T$.} \label{s31}
\end{figure}

From MFM (Magnetic Force Microscopy) measurements we know that higher
currents decrease the magnetic phase contrast at the center of the
spot, indicating the vanishing of magnetic order in part of the
irradiated area \cite{esq05}. To test this behavior with the SQUID
the spots in sample~3 were produced with a $\sim 9$ times larger
proton current than for the spots produced in sample~2.
Figure~\ref{s31}, as Fig.~\ref{s11}(b), shows the temperature
dependence of the magnetic moment produced by the irradiation of 400
spots (fluence 24.3~nC/$\mu$m$^2$ = $1.52 \times 10^{19}~$cm$^{-2}$,
total charge 191~$\mu$C). In contrast to the effects of the spots
produced in sample~2, the irradiation in sample 3 triggers a
paramagnetic contribution that follows the Curie law plus a smaller,
constant ferromagnetic-like background. This small, last contribution
is recognized also in the s-shape curve at low fields obtained from
the difference between the hysteresis loops.  The results indicate
that it is not only the total implanted charge what determines the
ferromagnetic and/or the paramagnetic effect but also the proton
current appears to have an important role due to overheating effects.

\section{Conclusion}
We have performed high sensitive measurements with a commercial SQUID
with RSO option of the effects produced by proton irradiation in the
MeV energy range on HOPG samples. All proton irradiations produce two
magnetic contributions to the magnetic properties of HOPG. One
contribution depends clearly on temperature and in some cases follows
a Curie-like behavior. The second one can be attributed to ferro- or
ferrimagnetism. These contributions are supported by the measured
hysteresis and s-form of the loops as a function of magnetic field.
The details of these contributions depend not only on the total
implanted charge, the fluence but also on the proton current and the
beam geometry, i.e. broad beam irradiation or localized spots.
Largest ferromagnetic signals were obtained for localized spots
produced at relatively low proton currents ($< 1~$nA) and fluences $
\lesssim 1~$nC/$\mu$m$^2 (6.25 \times 10^{17}~$cm$^{-2}$) . The
saturation magnetic moment per micrometer spot is of the order of
$10^{-10}~$emu. The absolute values of the observed effects are of
the same order as those published in the original work of
Ref.~\cite{pabloprl03}. The observed decrease of the effects at large
fluences and proton currents agrees with the decrease of the
(magnetic) phase contrast with fluence and current observed in MFM
measurements on similar spots produced in HOPG surfaces \cite{esq05}.

We note that both, SQUID and MFM measurements of the magnetic
behavior of micrometer size spots indicate that upon irradiation
conditions they may not behave ferromagnetically with a finite
remanent magnetic moment at zero field. In this case the
characterization of the spots by means of MFM under zero field
conditions is difficult since no significant contrast difference will
be measured after a change of the tip polarization or after applying
a magnetic field on the spots.

Three experimental methods were used to check the amount of magnetic
impurities in similar HOPG samples from the same company. The results
indicate that the total amount of magnetic impurities is much below
that needed to understand quantitatively the measured ferromagnetism
after irradiation. With the implementation of a special sample holder
for SQUID measurements and irradiation runs, we could increase the
reliability of the SQUID results and rule out the influence of
artifacts due to handling on the sample and its holder.

On the origin of the magnetic order in proton irradiated carbon we
note the following. The overall results suggest that carbon defects
(e.g. adatoms or vacancies) should play a role in the magnetic order
observed. According to recent models \cite{lehtinen04} hydrogen at
carbon defects (H-vacancy and -adatom complexes) may also contribute
triggering or even enhancing a local magnetic moment. Regarding the
role of hydrogen and taking into account that HOPG samples have a
substantial amount of hydrogen at the first micrometer from the
surface \cite{rei06}, one can speculate that not the implanted charge
but the dissociation of molecular hydrogen (already in the sample)
produced by the proton collisions may be important for the magnetic
order, since in this case single hydrogen atoms may be more effective
to bond at the magnetic sensitive defects.

An independent support to the SQUID and MFM results on irradiated
carbon samples is provided by x-ray magnetic circular dichroism
measurements on spots produced in 200~nm thick carbon films
\cite{ohldag}. These results also suggest that it is probably not the
implanted hydrogen but, {\em if at all}, the one already in the
sample of importance.

We recommend that before selecting a particular irradiation, taking
into account several of the irradiation parameters discussed in this
work, the reproducibility as well as the resolution limits of the
used SQUID should be checked in order to estimate the minimum
irradiation and sample requirements.

\end{document}